\documentstyle[prl,multicol,aps,epsf]{revtex}
\begin{document}

\title{
Nesting symmetries and diffusion in disordered $d$-wave 
superconductors} 

\author{ 
A.\ G.\ Yashenkin$^{1,2,\dagger}$,
W.\ A.\ Atkinson$^{1,3}$,
I.\ V.\ Gornyi$^{4,\ddagger}$, 
P.\ J.\ Hirschfeld$^1$, and 
D.\ V.\ Khveshchenko$^{2}$ 
}

\address{
$^1$ Department of Physics, University of Florida, PO Box 118440,
Gainesville FL 32611 \\ 
$^2$ Department of Physics and Astronomy, University of North 
Carolina, Chapel Hill, NC 27599 \\
$^3$ Department of Physics, Southern Illinois
University, Carbondale, IL 62901-4401 \\
$^4$ Institut f\"ur Nanotechnologie, Forschungszentrum 
Karlsruhe, 76021 Karlsruhe, Germany}
\date{\today} 

\maketitle

\begin{abstract}
The low-energy density of states (DOS) of disordered 2D $d$-wave 
superconductors is extremely sensitive to details of both the disorder 
model and the electronic band structure. Using diagrammatic methods 
and numerical solutions of the Bogoliubov-de Gennes equations, we show 
that the physical origin of this sensitivity is the existence of a 
novel diffusive mode with momentum close to $(\pi,\pi)$ which is 
gapless only in systems with a global nesting symmetry. We find that 
in generic situations, the DOS vanishes at the Fermi level. However, 
proximity to the highly symmetric case may nevertheless lead to 
observable non-monotonic behavior of the DOS in the cuprates.
\end{abstract}



\begin{multicols}{2}
\narrowtext

\textit{Introduction.}
An understanding of the quasiparticle (QP) excitations in the $d$-wave 
superconducting state of the high-$T_c$ superconductors is essential 
for the elucidation of transport properties, for determining how the 
ground state deviates from the BCS model, and for describing the 
instability of the lightly doped antiferromagnetic state to 
superconductivity. It has been known for some time that the influence 
of disorder on the QP states is quite different from ordinary 
superconductors, in part due to the gap symmetry and in part due to 
low dimensionality. Nersesyan {\it et  al}. have shown that these two 
features conspire to introduce logarithmic divergences in all orders 
of the perturbation theory \cite{NTW}. Since then, several groups have 
attempted nonperturbative treatments of the ``2D dirty $d$-wave 
problem'', arriving at a surprisingly diverse set of results. The 
proposed scenarios predict vanishing\cite{NTW,Fisher,Bocquet},
constant\cite{Ziegler,AltHuc}, and divergent \cite{PepinLee,mudry} 
density of states (DOS) $\nu(\epsilon)$ as $\epsilon\to 0$ (energies 
are measured from the Fermi level) for apparently similar models. 
Recently, two of the authors argued \cite{AHMZ} on the basis of 
numerical studies that the $d$-wave superconductor is fundamentally 
sensitive to ``details" of disorder, as well as to certain symmetries 
of the normal state Hamiltonian. While this approach was successful in 
unifying the various analytical treatments, it failed to provide a 
physical explanation of the origin of this lack of robustness.

In this work, we combine perturbative analytical and numerical
calculations with the intent of clarifying the physics of the various 
asymptotic results for the DOS of noninteracting QPs. We show that the 
generic result is $\nu(\epsilon\to 0)\to 0$, but that both the 
constant and divergent DOS can be obtained in the presence of global 
``nesting'' symmetries (GNSs) which produce additional diffusive modes 
with momenta close to ${\bf Q}=(\pi,\pi)$ \cite{nakh}. These symmetries 
drastically change the low-energy DOS and are explicitly obeyed by 
models considered in Refs. \cite{Ziegler} and \cite{PepinLee} where the 
constant and the divergent $\nu(\epsilon\to 0)$ have been found. The 
GNS relations necessary for the $\pi$-modes do not occur explicitly in 
real cuprates, but we demonstrate that even proximity to global nesting 
has observable consequences on thermodynamic properties.

\textit{Models of pure $d$-wave superconductor.}
The weak-localization (WL) calculations will be compared directly with 
exact solutions of the Bogoliubov-de Gennes equations on finite 
tight-binding lattices, details of which have been published elsewhere
\cite{AHMZ,AHM}. We investigate three different model band structures, 
each having specific symmetry properties. The nearest neighbor (N1) 
model with normal-state dispersion $\xi_{\bf k} = -2t(c_x + c_y)-\mu$,
where $t$ is the hopping matrix element and $\mu$ is the chemical 
potential, is the most extensively studied. 
Here $c_{x,y}=\cos(k_{x,y}a)$ and $a$ is the lattice constant.
We also consider a model including second-neighbor (N2) hopping,
$\xi_{\bf k} = -2t(c_x + c_y) -4t^\prime c_x c_y -\mu$   
and a peculiar third-neighbor (N3) hopping model 
$\xi_{\bf k} = -4 t^{\prime\prime} (c_{x}^2 + c_{y}^2 -1) -\mu$.
In all cases, the $d$-wave Hamiltonian assumes pairing 
between nearest neighbour sites, 
$\Delta_{\bf k} = \Delta_0 (c_x - c_y)$.
The lines of zeros for $\Delta_{\bf k}$ intersect the Fermi surface 
$\xi_{\bf k}=0$ at four symmetric nodal points for both the N1 and 
N2 models, and at eight points for the N3 model. In the vicinity of 
these points we may make an expansion of the QP spectrum 
$\varepsilon_{\bf k} \simeq [({\bf v}_{F}  \tilde{\bf k})^2 + 
({\bf v}_{g} \tilde{\bf k})^2 ]^{1/2}$, 
where ${\bf v}_{F} = (\partial \xi_{{\bf k}} / 
\partial {\bf k})_{{\bf k}_n}$,
${\bf v}_{g} = (\partial \Delta_{{\bf k}} / 
\partial {\bf k})_{{\bf
k}_n}$ are the Fermi and ``gap'' velocities, and $\tilde{\bf k}$ 
is the momentum measured from the node at ${\bf k}_n$. Gapless
nodal excitations determine the low-energy behavior of the clean 
DOS, $\nu_{cl} (\epsilon\to 0) = \alpha \epsilon $, where 
$\alpha =N/(2\pi v_{F} v_{g})$ and $N$ is the number of nodes.

\textit{Self-consistent T-matrix approximation.}
In sharp contrast with a normal metal, the DOS in $d$-wave 
superconductors is strongly affected by disorder. The commonly 
used self-consistent T-matrix approximation (SCTMA) yields finite 
DOS at the Fermi level, $\nu_0 = \frac{2}{\pi} \alpha \gamma l$ 
(see, e.g.,\ \cite{DL}). Here $\gamma$ is the impurity-induced 
relaxation rate of nodal excitations, $l=\ln (\Lambda/\gamma)$, 
and $\Lambda \sim (v_{F} v_{g})^{1/2}a^{-1}$ is the high-energy 
cut-off in the nodal expansion. 

\begin{figure}[tb]
\epsfxsize .9\columnwidth
\epsffile{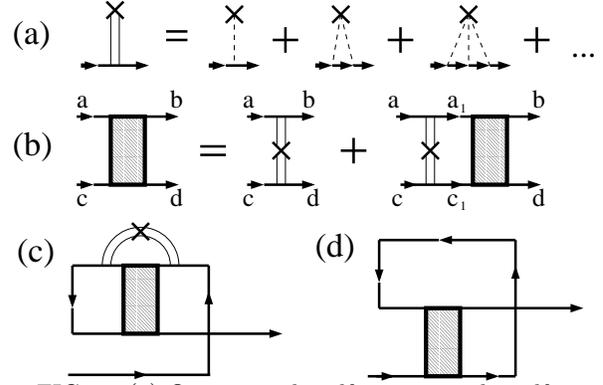}
\caption{(a) Quasiparticle  self-energy in the self-consistent 
T-matrix approximation; (b) Diagramatic equation for the 
Cooperon ${\cal C}$ (the direction of the arrows in the 
bottom line should be reversed for the diffuson ${\cal D}$); 
(c) diffuson and (d) Cooperon contributions to DOS. Solid 
lines are quasiparticles, and dashed lines indicate impurity 
potentials $U$.}
\label{fig:diagrams}
\end{figure}

Let us outline first the SCTMA for a general asymmetric band. 
Assuming that both the time-reversal and spin-rotational symmetries 
are preserved, consider a random distribution of point-like 
spinless scatterers of arbitrary strength $U$ and concentration 
$n_i$.  The self-consistent T-matrix 
$\hat{T}(\tilde\epsilon) = \sum_i T_i (\tilde\epsilon)\hat{\tau}_i$,
depicted in Fig.~\ref{fig:diagrams}(a), has two Nambu components
given by 
$T_3 (\tilde\epsilon ) \pm T_0 (\tilde\epsilon) = 
[c \mp g_0 (\tilde\epsilon)]^{-1}$,
with $c = U^{-1} - g_3(\epsilon=0)$ and
$\tilde\epsilon = \epsilon - \Sigma_0$.
Here $\hat{\tau}_{1,2,3}$ are the Pauli matrices and $\hat{\tau}_0$ 
is the unity matrix, $\Sigma_0 = n_i T_0$, and the chemical 
potential shift is $\delta \mu = \Sigma_3 = n_i \, T_3$. The 
quantities $g_{0,3}$ are the Nambu components of the integrated 
Green's function
$\hat{g} (\tilde\epsilon) = 
\sum_{\bf k} \hat{G}_{{\bf k}} (\tilde\epsilon)$, where
$\hat{G}_{{\bf k}} (\tilde\epsilon) = (\tilde\epsilon
\hat \tau_0 +\Delta_{\bf k}\hat \tau_1 +
\xi_{\bf k}\hat\tau_3)/(\tilde\epsilon^2 - \varepsilon^2_{\bf k})$.
The above, closed system of equations apply to both the retarded (R) 
and advanced (A) channels, and it is understood that $\epsilon$ 
lies in the upper or lower half-plane as appropriate. We remark that 
the quantity $g_3(\epsilon = 0)$ is real and vanishes for a 
perfectly symmetric band; it therefore cannot be estimated in a 
nodal expansion. At the level of SCTMA, $g_3$ just renormalizes 
the effective scattering potential with $U \to c^{-1}$.
The Born and unitarity limits correspond to $c \gg |g_0|$ 
and $c \to 0$ respectively. The QP self-energies (QPSEs) are 
non-singular at small frequencies $\epsilon < \gamma$ where they
obtain the form 
$\Sigma^{R(A)}_0(\epsilon) = \lambda \epsilon \mp i \gamma$.
Here the relaxation rate $\gamma$ is the solution 
(cf.\ \cite{DL,lee}) of $\gamma = i n_i T_0(\epsilon =0)$,
and the mass renormalization satisfies
$\lambda = n_i \partial_\epsilon  
T_0(\tilde\epsilon) |_{\epsilon=0}$.

\textit{Analysis of diffusion modes.} 
Since the SCTMA is exact only for a single-impurity problem, $\nu_0$ 
is subject to further corrections caused by multiple impurity 
scattering which result in new low-energy regimes 
\cite{NTW,Fisher,Bocquet,Ziegler,AltHuc,PepinLee}. 
We explore these regimes by calculating the WL correction to the 
SCTMA result in the diffusion approximation. We observe that the 
first-order correction predicts many aspects of the low-energy 
behavior and qualitatively agrees with numerical calculations. In 
particular, the sign of the correction can be associated with the 
further tendency for $\nu (\epsilon\to 0)$: to diverge (positive 
sign), to vanish (negative sign), or to saturate (correction is 
absent). 

Consider the diffuson 
$\hat {\cal D}_{\bf q} (\epsilon,\epsilon^\prime)$ and Cooperon
$\hat {\cal C}_{\bf q}(\epsilon,\epsilon^\prime)$ ladder 
diagrams depicted in Fig.~\ref{fig:diagrams}(b). The 2-particle 
irreducible vertex 
$\hat{\cal I}_{\tilde\epsilon, \tilde\epsilon^\prime} = 
n_i \hat T(\epsilon) \otimes \hat T(\epsilon^\prime)$.  
The solution of the matrix equations for $\hat {\cal D}$ and 
$\hat {\cal C}$ is simplified if we make the decomposition 
${\cal X}_{abcd} = 
\frac{1}{2} \sum_{ij} {\cal X}^{ij} {\tau_i}_{ad} {\tau_j}_{cb}$ 
for ${\cal X} = {\cal I},\, {\cal D,\, C}$.
In this basis, the ladder diagrams reduce to a $4\times 4$ matrix 
equation. Furthermore, it is easy to show  that 
${\cal C}^{ij}_{\bf q}(\epsilon,\epsilon^\prime)
= {\cal D}^{ij}_{\bf q}(\epsilon,\epsilon^\prime)$, 
and that only the diagonal components of the diffusive modes 
may be singular, 
\begin{equation}
{\cal D}^{ii}_{\bf q}(\epsilon, \epsilon^{\prime})
=\frac{{\cal I}^{ii}_{\epsilon, \epsilon^{\prime}}}
{1-{1\over{2}}{\cal I}^{ii}_{\epsilon,\epsilon^{\prime}}
\sum_{\bf k}{\rm Tr}
[ \hat{G}_{\bf k+q}(\tilde\epsilon)\hat{\tau}_i \hat{G}_{\bf k}
(\tilde\epsilon^{\prime}) \hat{\tau}_i]},
\label{formD}
\end{equation}
where ${\cal I}^{ij}_{\epsilon, \epsilon^{\prime}}=\frac{1}{2}
n_i{\rm Tr}[{\hat T}(\tilde\epsilon){\hat \tau}_i{\hat T}
(\tilde\epsilon^{\prime}){\hat \tau}_j]$.
Besides the singular contributions, diffusive ladders contain 
the regular ones. For superconductors, the need to sum up the 
nonsingular ladders has been pointed out in 
Ref.\ \cite{AltZirn1}.

In the RA channel, ${\cal D}^{00}$ and ${\cal C}^{00}$ have 
diffusive poles at small $\bf q$ and 
$\epsilon - \epsilon^{\prime}$,
$${{\cal D}^{00}_{\bf q}(\epsilon, \epsilon^{\prime}) }^{RA}
={{\cal C}^{00}_{{\bf q}}(\epsilon, \epsilon^{\prime}) }^{RA}
= \frac{4 \gamma^2}
{\pi \nu_0} \frac{1}{D_0 q^2 - i (\epsilon - \epsilon^{\prime})}.
$$

For diffusons and Cooperons, ${\bf q}$ is the difference and sum
of momenta of QPs in the top and bottom lines in 
Fig.\ \ref{fig:diagrams}(b) 
respectively, and the diffusion coefficient $D_0$ is found to be 
$D_0 =\overline v^2/(2 \gamma l)$ with 
$\overline v^2= \frac12 (v_F^2+v_g^2)$.
The logarithmic factor $l$ in the denominator of $D_0$ 
originates from renormalization of the real part of the QPSE. 
Upon this renormalization, $D_0$ and $\nu_0$ obey the Einstein 
relation $D_0 \nu_0 = \sigma_s / s^2 = 
\frac{1}{\pi} \alpha \overline v^2$, 
where $\sigma_s$ is the universal value of the Drude spin 
conductance \cite{Fisher,DL}, and $s=\frac{1}{2}$ is the spin 
of electron.
 
Due to the intrinsic particle-hole symmetry of a
superconductor,
$\hat{\tau}_2 \hat{G}_{\bf k}(-\tilde\epsilon)\hat{\tau}_2
= -\hat{G}_{\bf k}(\tilde\epsilon)$, 
diffusive modes also exist in the RR and AA channels 
\cite{AltZirn1}. These anomalous modes appear in 
${\cal D}^{22}$ and ${\cal C}^{22}$  when 
$\epsilon, \, \epsilon^{\prime} < \gamma$,
\begin{equation}
{{\cal D}^{22}_{\bf q} (\epsilon, \epsilon^{\prime}) }^{RR}=
{{\cal C}^{22}_{\bf q} (\epsilon, \epsilon^{\prime}) }^{RR}=
-{{\cal D}^{00}_{\bf q} (\epsilon, -\epsilon^{\prime}) }^{RA}.
\label{eq:diff22}
\end{equation}
They are always gapless due to the fact that the spin and the energy 
are conserved quantities in the problem.
 
{\it WL correction in generic case.}
In the absence of special symmetries, Eq.~(\ref{eq:diff22}) 
gives the only diffusive modes which contribute to the DOS. 
The first-order WL correction can be written in the form
\begin{equation}
\delta \nu (\epsilon) = -\frac{1}{\pi}{\rm Im} \sum_{{\bf k}} 
{\rm Tr} [ \hat{G}^{R}_{{\bf k}}(\tilde\epsilon) 
\hat\Sigma^{R}_{{\bf k}}(\tilde\epsilon) 
\hat{G}^{R}_{{\bf k}}(\tilde\epsilon)].
\end{equation}
The skeleton diagrams for the self-energy
$ \hat\Sigma^{R}_{{\bf k}} (\tilde\epsilon)$ 
are depicted in Figs.~\ref{fig:diagrams}(c) and (d), with 
solid grey blocks denoting diffuson or Cooperon propagators.
Diagram \ref{fig:diagrams}(c) -- the diffuson with one leg 
decorated by an irreducible vertex, as well as a similar 
diagram with the other leg decorated -- vanishes, so the 
leading order contribution is the Cooperon diagram, 
\ref{fig:diagrams}(d), which yields the negative logarithmic 
correction (as found in \cite{Fisher,KYG})
\begin{equation}
\delta \nu (\epsilon) = - \frac{1}{4 \pi^2 D_0 l} 
\ln \frac{\gamma}{2\epsilon}. 
\label{eq:dos1}
\end{equation}
This holds for generic bands and arbitrary strength of disorder. 
Since $D_0 l = \overline v^2 /(2 \gamma)$, the high-energy 
cut-off $\Lambda$ enters Eq.\ (\ref{eq:dos1}) via the scattering 
rate $\gamma$ only. Furthermore, the insertion of nonsingular 
ladders into diagrams Fig.\ \ref{fig:diagrams}(c)-(d) in all 
possible ways leads to ballistic renormalization of the 
parameter $D_0$ in Eq.\ (\ref{eq:dos1}). After the 
renormalization, $D_0 \to D$, this expression is the 
complete singular contribution to the DOS in first order of 
perturbation theory in the inverse spin conductance 
$\sigma_{s}^{-1}$.

\begin{figure}[tb]
\epsfxsize .9\columnwidth
\epsffile{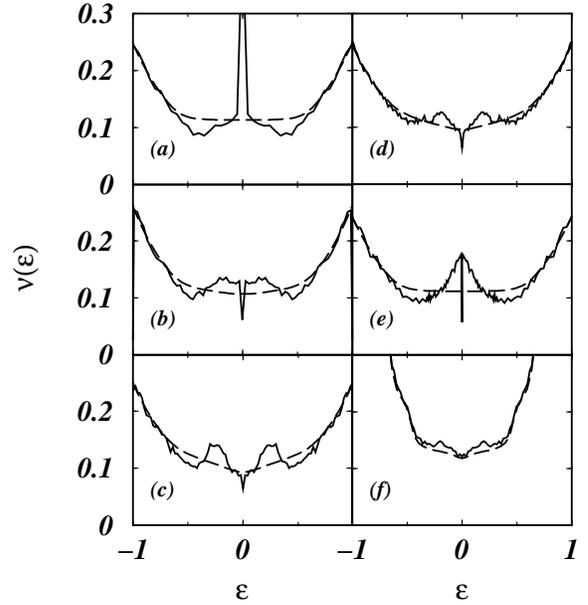}
\caption{Numerical results for $n_i=0.06$ and $\Delta_0 = 0.8$.  
Solid curves are solutions of BdG equations on a $30\times 30$ 
lattice [except for (f), which is on a $45\times 45$ lattice], 
dashed curves are SCTMA results.  Curves are N1 with 
$\{\mu,U,c\}=$ (a) $\{0,10^4,10^{-4}\}$, (b) $\{0.6,10^4,0.07\}$, 
(c) $\{0,10,0.10\}$; N2 with (d) $\{0,10^4,0.099\}$, (e) 
$\{0,-11.675,-8\times 10^{-5}\}$, and N3 with (f) $\{0,10,0.15\}$.  
Energies are measured in units of $t$ for N1 and N2 models, and
$t^{\prime\prime}$ for the N3 model. The value of $t^{\prime}$ 
for N2 model is chosen to be 0.25.}
\label{fig:numerics}
\end{figure}

The supression of the DOS relative to $\nu_0$ at low energies
indicated by Eq.\ (\ref{eq:dos1}) is seen 
in nearly all published numerical work \cite{AHMZ,AHM,tinran}, 
as well in most cases presented in Fig.~\ref{fig:numerics}. A 
glance at Fig.~\ref{fig:numerics}(b)-(e) shows that, although  
$\nu(\epsilon\to 0)\to 0$ in these cases, there is typically more 
structure than is contained in Eq.~(\ref{eq:dos1}). In order to 
understand this better, we consider the effects of certain 
nesting symmetries.

\textit{Global nesting symmetry $\hat{\tau}_2$.} 
The above Goldstone modes Eq.~(\ref{eq:diff22}) occur for
arbitrary renormalized potential $c^{-1}$. As it is seen from 
Eq. (\ref{formD}), additional pseudo-Goldstone modes appear 
in the unitarity limit for systems which satisfy a GNS,
\begin{equation}
{\hat{\tau_2}} \hat{G}_{{\bf k}} {\hat{\tau_2}} = 
\hat{G}_{{\bf k +{\bf Q}}}.
\label{eq:symmetry}
\end{equation}
When Eq.\ (\ref{eq:symmetry}) is exact {\it and} $c=0$ (e.g.,\ 
in the $\mu=0$, $U=\infty$ N1 model studied in  
\cite{PepinLee}) the propagators ${\cal D}_{{\bf q}}$ and 
${\cal C}_{{\bf q}}$ 
have additional poles at momenta close to 
${\bf Q}=(\pi,\pi)$.
These modes are related to the usual ones by 
\begin{eqnarray}
{{\cal D}^{22}_{{\bf Q}+{\bf q}} 
(\epsilon, \epsilon^{\prime})}^{RA} =&
{{\cal C}^{22}_{{\bf Q}+{\bf q}} 
(\epsilon, \epsilon^{\prime})}^{RA} =& 
{{\cal D}^{00}_{\bf q}
(\epsilon, \epsilon^{\prime})}^{RA},  
\nonumber \\
{{\cal D}^{00}_{{\bf Q}+{\bf q}}
( \epsilon, \epsilon^{\prime})}^{RR} =& 
{{\cal C}^{00}_{{\bf Q}+{\bf q}}
( \epsilon, \epsilon^{\prime})}^{RR} =& 
{{\cal D}^{22}_{\bf q}( \epsilon, \epsilon^{\prime})}^{RR}.  
\nonumber 
\end{eqnarray}
However, these 
modes are gapped by any distortion of the 
band which destroys the nesting symmetry or by the choice of 
nonunitarity potential ($c \neq 0$),
as we now show.

Firstly, we note that the N1 model at $\mu, c=0$ satisfies 
the GNS conditions {\it exactly}, so the $\pi$-modes are 
gapless in this case. Deviation from the half-filling 
($\mu \neq 0$) creates a gap which can be estimated in the 
nodal approximation as $\delta \approx 2\mu^2 /(\gamma l)$, 
whereas deviation from unitarity ($c \neq 0$) yields 
$\delta \approx \gamma (4 c / \pi \nu_0)^2$. Secondly, 
the N2 model has nodes at 
$(\pm \pi/2, \pm \pi/2)$ as $\mu=0$, 
and the symmetry relation (\ref{eq:symmetry}) is 
approximately satisfied near the nodal points.  
However, this is not the case in the entire Brillouin zone 
where the $t^{\prime}$-term evidently violates 
Eq.\ (\ref{eq:symmetry}). As a result, the momentum regions 
located far away from the nodal points (these are momenta 
responsible for the global band asymmetry) contribute to a 
gap $\delta \propto t^{\prime 2}$ in  
N2 model even at $c=0$. We stress that this contribution 
cannot be estimated in the nodal approximation. Therefore, 
we conclude that the GNS relation (\ref{eq:symmetry}) must hold 
througout the Brillouin zone, and not just near the 
gap 
\noindent
nodes; both deviations from Eq.~(\ref{eq:symmetry}) and 
deviations from  unitarity gap the $\pi$-modes.

\textit{ WL correction in $\hat{\tau}_2$ case}.
The $\pi$-mode Cooperon makes a logarithmic contribution to 
$\delta \nu(\epsilon)$ of equal magnitude to Eq.~(\ref{eq:dos1}) 
but -- because it appears with different Nambu components -- of 
opposite sign. The diffuson in Fig.~\ref{fig:diagrams}(c) makes 
a positive logarithmic contribution so the overall tendency is 
for $\delta \nu(\epsilon)$ to be positive. Thus, the additional 
symmetry (\ref{eq:symmetry}) changes the asymptotic behavior 
from vanishing to divergent. If the $\pi$-modes are gapped 
but $\delta < \gamma$, then we identify two regimes: 
$\epsilon > \delta/2$ where gapless ordinary modes and the gapped
$\pi$-modes contribute on equal footing such that the total 
correction increases, and $\epsilon < \delta/2$ where the 
$\pi$-mode contribution saturates whereas the ordinary one 
continue to suppress the DOS. Generally, the sum of these two 
terms can be written as follows
\begin{equation}
\delta \nu (\epsilon) = \frac{1}{4 \pi^2 D_0 l}
\left( - \ln \frac{\gamma}{2\epsilon} + 5 \ln
\frac{\gamma}{\sqrt{4\epsilon^2 + \delta^2}} \right).
\label{eq:dos2}
\end{equation}
Numerical calculations \cite{AHMZ,AHM,tinran}, 
as well as nonperturbative analytical work on the 
half-filled N1 model \cite{PepinLee} support this result. 
Fig.~\ref{fig:numerics}(a) shows a clear peak at $\epsilon=0$ for 
the N1 model with $\mu, c= 0$. In Figs.~\ref{fig:numerics}(b)-(e) 
the $\pi$-modes are gapped in three different ways. In all cases, 
the qualitative behavior is the same: there is an initial upturn 
starting at $\epsilon \simeq \gamma$, which then crosses over to 
a downturn. As we move farther from the symmetric unitarity limit, 
either by increasing $\mu$, decreasing $U$ or increasing 
second-neighbour hopping, the peaks are shifted to higher energy, 
then become less pronounced, and after that disappear. In (b)-(e),
the shape of the curves is qualitatively what one might expect 
based on Eq.~(\ref{eq:dos2}). While this expression regularly 
overestimates the position of the maxima predicted as 
$\epsilon \sim \delta/4$ (cf.\ Fig.~\ref{fig:numerics}), 
it is clear that the qualitative 
explanation based on the first-order perturbation correction 
could not pretend to describe all the details of the numerical  
experiment. We mention finally that the numerics 
[see Fig.~\ref{fig:numerics}(e)] and our WL calculation are 
in agreement that it is not possible to produce a divergence 
by means of fine-tuning parameters in the non-nested case, as 
suggested in Ref.\ \cite{mudry}. 

\textit{ Models with other nesting symmetries}.
There is a second possible nesting condition,
\begin{equation}
{\hat{\tau_3}} \hat{G}_{{\bf k}} {\hat{\tau_3}} = \hat{G}_{{\bf
k}+{\bf Q}},
\label{eq:symmetry2}
\end{equation}
which leads to a non-generic DOS. A model (N3) with this 
symmetry which yields an exact solution was recently discussed 
in \cite{Ziegler}. In this instance it was found, at least for 
Lorentzian disorder, 
that $\nu(\epsilon)$ remains finite as $\epsilon \to 0$.  
A similar result is seen in the WL 
calculations. While the  0-modes are unaffected by
Eq.~(\ref{eq:symmetry2}), the $\pi$-modes change drastically.  
Divergences now appear in ${\cal D}^{11}$ and ${\cal D}^{33}$ 
with
\begin{eqnarray*}
{{\cal D}^{33}_{{\bf Q}+{\bf q}}(\epsilon,\epsilon^\prime)}^{RA}=&
{{\cal C}^{33}_{{\bf Q}+{\bf q}}(\epsilon,\epsilon^\prime)}^{RA}=&
{{\cal D}^{00}_{\bf q}(\epsilon,\epsilon^\prime)}^{RA},\\
{{\cal D}^{11}_{{\bf Q}+{\bf q}}(\epsilon,\epsilon^\prime)}^{RR}=&
{{\cal C}^{11}_{{\bf Q}+{\bf q}}(\epsilon,\epsilon^\prime)}^{RR}=&
{{\cal D}^{22}_{\bf q}(\epsilon,\epsilon^\prime)}^{RR}.
\end{eqnarray*}
In this case, the diffusive modes contribute for {\it arbitrary} $c$.
Calculating the WL correction to the DOS we observe, as before, the 
exact cancellation between $0$- and $\pi$-Cooperons. However, the
contribution of $\pi$-diffusons [Fig.\ \ref{fig:diagrams}(c)] now 
also sums to zero, and the total first-order WL correction to DOS is 
$\delta \nu (\epsilon) = 0$. 

The N3 model explicitly satisfies the symmetry (\ref{eq:symmetry2}) 
for arbitrary $\mu$, and a typical case is shown in 
Fig.\ \ref{fig:numerics}(f). As expected, the DOS is finite at 
$\epsilon =0$, and agrees closely with the SCTMA result. This resembles the 
situation in a noninteracting normal metal \cite{Bocquet}, where the 
disorder by itself does not affect the DOS \cite{AA}. 
On the other hand, this model exhibits a Meissner effect. 
It is clear, however, 
that Eq.\ (\ref{eq:symmetry2}) describes a specific situation which 
is probably  unrelated to the real cuprates \cite{AHMZ}.
We note that the approach of Ref. \cite{mudry} fails to predict the
characteristic behaviors of the $\hat\tau_2$ and $\hat\tau_3$
models identified here.

In order to complete our analysis of various extra symmetry
possibilities, let us mention the purely artificial case:
${\hat{\tau_1}} \hat{G}_{{\bf k}} {\hat{\tau_1}} = 
\hat{G}_{{\bf k}+{\bf Q}}$.  
Then the gapless $\pi$-modes appear in the unitarity regime only, 
for which case we predict the constant $\nu (\varepsilon \to 0)$, 
since $0$- and $\pi$-Cooperons cancel each other whereas the 
diffuson contributions sum to zero. 
 
In conclusion, we investigated the low-energy density of states 
in a disordered 2D $d$-wave superconductor and argued that the 
generic feature of DOS is its low-energy suppression. However, due 
to proximity to perfect nesting, 
pseudo-Goldstone diffusive mode with momenta close to 
$(\pi,\pi)$ 
gives
rise to an energy
dependence of DOS which can be strongly non-monotonic. The
physics of this novel mode appears to account for the differences
between many recent nonperturbative treatments and explains
numerical results. The effects we predict should be visible
in low-temperature specific heat experiments on disordered cuprates.
In layered compounds with weak interlayer coupling $t_{\perp}$, WL
corrections to the DOS appear for $t_{\perp} < \gamma $, but saturate 
for $\epsilon < t_{\perp}^{2}/\gamma $. Thus the criterion for a 
pronounced energy window in which WL corrections occur is
$t_{\perp}^{2}/\gamma <\epsilon < \gamma $. This regime should be 
accessible in underdoped highly disordered  materials.
 
The authors thank  
D. Diakonov, I. Gruzberg, I. Lerner, C. Mudry, C. P\'epin, 
S. Sachdev, and A. Shytov for useful discussions. 
AGY, IVG, and DVK are grateful to NORDITA for hospitality.
This work was supported by 
NSF Grants 
DMR-9974396 and INT-9815833 (WAA and PJH), 
DMR-9703388 (AGY),
DMR-0071362 (DVK),
by RFBR (IVG), 
and in part by INTAS. 

\end{multicols}
\end{document}